\documentclass[12pt, journal, onecolumn, draftclsnofoot]{IEEEtran}

\usepackage{mathptmx}
\usepackage{amsmath}
\usepackage{graphicx}
\usepackage{subcaption}
\DeclareMathOperator{\mean}{mean}

\usepackage{hyperref}
\usepackage{listings}
\newcommand{\radar}{{\sc{radar}}}
\newcommand{\radarurl}{\url{https://ucl-badass.github.io/radar/}}

\newcommand*{\radarfont}{\fontfamily{cmr}\selectfont}
\newcommand{\radvar}[1]{{\radarfont #1}}

\newcommand{\NB}{\mathit{NB}}
\newcommand{\EVTPI}{\radvar{EVTPI}}
\newcommand{\EVPPI}{\radvar{EVPPI}}
\newcommand{\Max}{\radvar{\textbf{Max}}} 
\newcommand{\EV}{\radvar{\textbf{EV}}}

\newcommand{\matrixVar}[1]{{\overline{\overline{#1}}}}
\newcommand{\bigO}{\mathcal{O}}

\begin{document}

\title{Scalability Analysis of the RADAR Decision Support Tool}

\author{\IEEEauthorblockN{Saheed A. Busari, Emmanuel Letier}\\
\IEEEauthorblockA{Department of Computer Science\\
University College London\\
London, United Kingdom\\
\{saheed.busari.13, e.letier\}@ucl.ac.uk}
}

\maketitle


This report presents a theoretical complexity analysis and empirical scalability analysis of the Requirements and Architecture Decision Analyser (\radar{}) described in a separate paper~\cite{radar-icse17}. 
We will refer to that paper as the \radar{} paper and assume the reader is familiar with its content. Further information can be found in the tool website (\radarurl).

\section{Complexity Analysis}

The analysis of a \radar{} model involves four main steps:

\begin{enumerate}
\item Generating the design space
\item Simulating all solutions in the design space
\item Shortlisting the Pareto-optimal solutions
\item Computing expected values of information over the shortlisted solutions
\end{enumerate}

The time and space complexities of these steps are summarised in Table \ref{table:TimeSpaceComplexity}.

\begin{table*}[t]
\centering\begin{tabular}{ l l l }
\hline
Analysis Step & Time Complexity & Space Complexity  \\ 
\hline

Generating the design space& $\bigO(m)$ & $\bigO(m)$\\

Simulating all solutions in the design space	 & $\bigO(|Obj| \times |DS| \times N \times m)$&$\bigO(|\textit{Obj}| \times |DS| \times N)$ \\
 
Shortlisting the Pareto-optimal solutions	&$\mathcal{O}({|DS|^2})$ &  $\mathcal{O}({|DS|^2})$ \\ 

Computing expected information value over the shortlisted solutions	&   $\bigO(N \times |S|)$& $\bigO(N \times |S|)$ \\
\hline
\end{tabular}	
\caption{Time and Space Complexity of RADAR analysis algorithms. $m$ is the  number of nodes in the model's AST, $\textit{Obj}$ is the model objectives,  $DS$ is the Design Space,  $N$ is the number of simulations and $S$ is the set of shortlisted solutions (i.e. the set of Pareto-Optimal solutions).}
\label{table:TimeSpaceComplexity}
\end{table*}

The first step, generating the design space, involves a single recursive traversal of the model's abstract syntax tree (AST). The time and space complexity of this step is thus $\bigO(m)$ where $m$ is the model length measured as number of nodes in the model's AST.

The second step is the most computationally expensive. It generates a matrix ${\textit{SimResult}}$ of dimension $|{\textit{DS}}| \times |\textit{Obj}|$ where ${\textit{DS}}$ is the model's design space and $\textit{Obj}$ is the set of model's objectives ($\textit{SimResult}[s, obj]$ denotes the simulated value of objective $\textit{obj}$ for solution $s$). Each cell in the matrix is computed by generating $N$ simulations of the objective's random variable. Each simulation involves a single recursive traversal of the model's AST and is thus $\bigO(m)$. Generating $N$ simulations for all objectives and all solutions thus has a time of $\bigO(|\textit{Obj}| \times |DS| \times N \times m)$.


The third step involves finding the Pareto-optimal solutions in the $\textit{SimResult}$ matrix generated in the second step. Our implementation involves comparing pairs of solutions and has a worst case complexity of $\bigO(|DS|^2)$. We could also use a faster algorithm with a complexity of $\bigO(|DS| log |DS|)$ when $|Obj| \leq 3$  and $\bigO(|DS| (log |DS|)^{|Obj| - 2})$ when $|Obj| \geq 4$~\cite{Kung75} but since, as will be seen later, the running time of finding the Pareto-optimal solutions is small compared to simulation time, such optimisation would have no visible effect on the total analysis time.

The fourth step involves computing the expected value of total perfect information ($\EVTPI$) and the expected value of partial perfect information ($EVPPI$). 
$\EVTPI$ and $\EVPPI$ are always evaluated with respect to a given objective, noted \Max{} \EV(\textit{NB}), and for a given set $S$ of alternative solutions. 
Our implementation takes $S$ to be the set of Pareto-optimal solutions shortlisted in step 3.
We estimate $\EVTPI$ using the classic formula:
\[
	\EVTPI = \underset{i:1..N}{\mean} \max_{j: 1..M}\matrixVar{\NB}[i,j] - \max_{j: 1..M} \underset{i:1..N}{\mean} \matrixVar{\NB}[i, j].
\]
where $\matrixVar{\textit{NB}}$ is the simulation matrix that contains simulations of \textit{NB} for each solution in $S$~\cite{letier-icse14}. The time complexity of such operation is $\bigO(N \times |S|)$. We compute $\EVPPI$ using a recent efficient algorithm that estimates $\EVPPI$ for a parameter $x$ from $\matrixVar{\textit{NB}}$ and and the vector $\overline{x}$ containing the simulations of parameter $x$~\cite{need-for-speed}. The complexity of this algorithm is also $\bigO(N \times |S|)$.

\section{Empirical Scalability Analysis}

In the main \radar{} paper~\cite{radar-icse17}, we report the application and running time of \radar{} analysis on four real-world examples. The largest model (the analysis of architecture decisions for an emergency response system) has a design space of 6912 solutions and takes 111 seconds to analyse (less than 2 minutes). In this section, we further evaluate \radar{}'s scalability by measuring its running time on larger synthetic models. 

We perform experiments to answer the following research questions:
\begin{itemize}
\item \textbf{RQ1:} What is \radar{}'s scalability with respect to the number of simulations?
\item \textbf{RQ2:} What is \radar{}'s scalability with respect to the size of the design space?
\item \textbf{RQ3:} What is \radar{}'s scalability with respect to the  number of objectives?
\item \textbf{RQ4:} What portion of time is spent on each of the four analysis steps?
\end{itemize}

To perform experiments answering these questions, we have implemented a synthetic model generator that generates random \radar{} models with a given number of objectives, decisions,  number of options per decisions and minimum number of model variables. The model generator can produce \radar{} models with or without decision dependencies. 
 

The model generator and all models generated for the experiments below are available from the tool's website (\radarurl).  All our experiments are run  on a machine running Linux with a four-core 2.6 GHz processor and 10GB RAM.

\subsection*{RQ1: What is RADAR's scalability with respect to the number of simulations}

\begin{figure*}
\centering
        \begin{subfigure}[b]{0.48\textwidth}
                \centering
                \includegraphics[width=\linewidth]{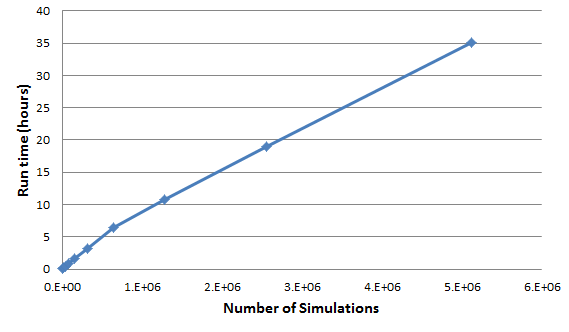}
                \label{fig:gull}
        \end{subfigure}
        \begin{subfigure}[b]{0.48\textwidth}
                \centering
                \includegraphics[width=\linewidth]{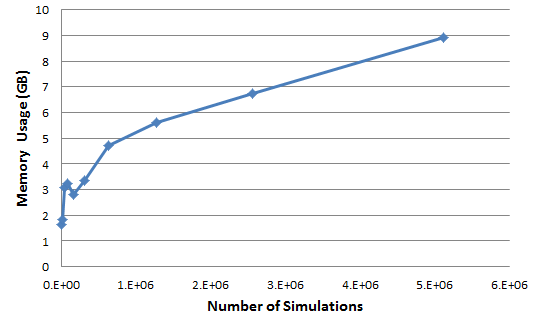}
               \label{fig:gull2}
        \end{subfigure}
        \caption{Total run-time (left) and memory usage (right) measured for doubling the number of simulations, $N$, from $10^4$ to $512\times10^4$. }
        \label{fig:RQ1-Simulations}
\end{figure*}

To evaluate how \radar{} run-time and memory usage increases as the number of simulation, $N$, increases, we have generated a synthetic model whose characteristics are similar to that of the emergency response system, i.e. it contains 2 objectives, 10 decisions, 3 options per decisions, and no decision dependencies. 
We have then measured the running times and memory consumption of analysing this model when doubling $N$ 10 times from $10^4$ to $512 \times 10^4$. The results are shown in Figure \ref{fig:RQ1-Simulations}. 
They indicate that the running time and memory usage increase linearly with $N$ as expected from the theoretical complexity analysis.

\subsection*{RQ2: What is RADAR's  scalability with respect to design space size}

To evaluate how \radar{} run-time and memory usage increases when the design space size increases, we have generated synthetic models by incrementally increasing the number of decisions and options per decisions until the resulting models could no longer be analysed in less than an hour. All models generated have 2 objectives and at least 100 model variables. The simulations are performed with $N = 10^4$.

\begin{figure*}
\centering
        \begin{subfigure}[b]{0.48\textwidth}
                \centering
                \includegraphics[width=\linewidth]{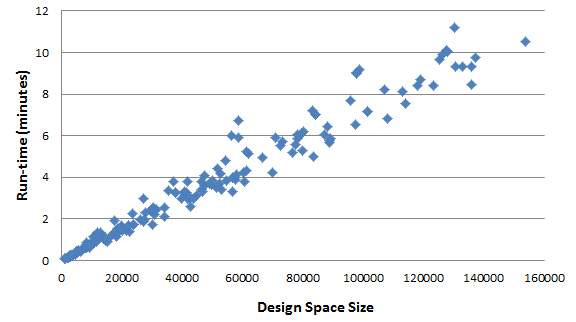}
                \label{fig:gull}
        \end{subfigure}
        \begin{subfigure}[b]{0.48\textwidth}
                \centering
                \includegraphics[width=\linewidth]{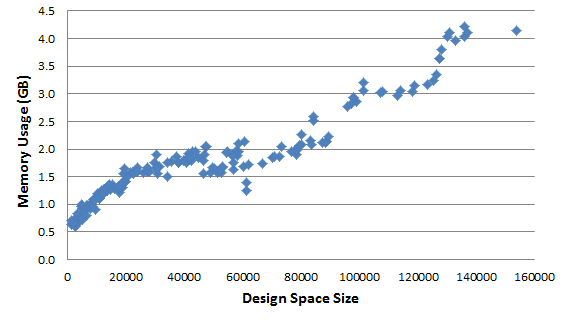}
               \label{fig:gull2}
        \end{subfigure}
        \caption{Total run-time (left) and memory usage (right) measured for 180 \radar{} models with different design space size. }
        \label{fig:RQ2-DesignSpace}
\end{figure*}

Figure \ref{fig:RQ2-DesignSpace} shows the result of this experiment. 
The largest model \radar{} was able to evaluate in less than one hour models has a design space of up to 153,751 solutions and includes 11 decisions with 7 options per decision. The figure also show that, as expected from the theoretical analysis, the run-time and memory usage increase roughly linearly with the size of the design space.

\subsection*{RQ3: What is RADAR's  scalability with respect to the number of objectives}

To evaluate how \radar{} run-time and memory usage increases when the number of objectives increases, we have generated synthetic models with 10 decisions, 3 options per decisions, and incrementally increased the number of objectives from 2 to 5 until the resulting models could no longer be analysed in less than an hour. The synthetic models generated do not have decision dependencies. The size of their design space is thus $3^{10} = 59,049$.

\begin{figure*}
\centering
        \begin{subfigure}[b]{0.48\textwidth}
                \centering
                \includegraphics[width=\linewidth]{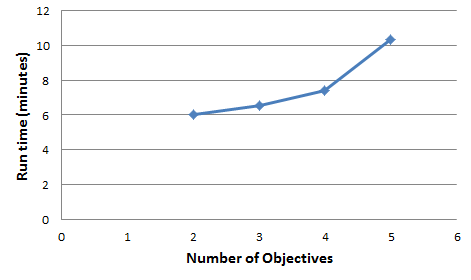}
                \label{fig:gull}
        \end{subfigure}
        \begin{subfigure}[b]{0.48\textwidth}
                \centering
                \includegraphics[width=\linewidth]{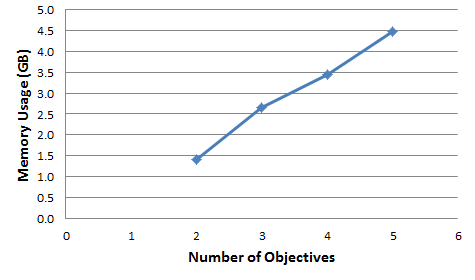}
               \label{fig:gull2}
        \end{subfigure}
        \caption{Total run-time (left) and memory usage (right) measured for 2,3,4 and 5 objectives. }
        \label{fig:RQ3-Objective}
\end{figure*}

Figure \ref{fig:RQ3-Objective} shows that on our synthetic models the run-time and memory usage increase roughly linearly with the number of objectives, as expected from the theoretical complexity analysis.

\subsection*{RQ4: What is the time spent and memory consumed by each analysis step}

For each synthetic model generated in the experiment to answer RQ2, we measured the fraction of time spent and memory used in each of the four analysis step: generating the design space, simulating the design space, shortlisting the Pareto-optimal solutions, and computing expected information value. Table \ref{table:TimeSPent} shows the average fraction of time for each analysis step over all synthetic models analysed in answer to RQ2 and RQ3. The table shows that the simulation of all solutions takes the largest portion of time (100\%) and memory consumption (96\%).

\begin{table}[t]
\centering
\begin{tabular}{l r r }
\hline
Algorithm Step & Average  \%  Total Time &  Average   \% Memory Usage\\ 
\hline
Generating the design space & 0 & 0 \\ 
Simulating all solutions in the design space& 100 & 96 \\ 
Shortlisting the Pareto-optimal solutions & 0  & 1   \\ 
Computing expected information value over the shortlisted solutions & 0  &  3\\ 
\hline
\end{tabular}
\caption{Real world \radar{} applications and their problem sizes.}
\label{table:TimeSPent}
\end{table}

\section{Conclusion}

We have shown both theoretically and empirically that \radar{}'s running time and memory usage increase linearly with the number of simulations, number of objectives and size of the design space. 
We also observed that \radar{}'s running time is almost entirely consumed by the exhaustive simulation of the design space.


The design space size is the critical factor limiting the scalability of \radar{}'s exhaustive simulation of the design space. 
Without decisions dependencies, the design space size increases exponentially with the number of decisions. This severely limits the class of decision problems \radar{} can currently analyse.
As a rule-of-thumb, \radar{}'s exhaustive search strategy would struggle solving problems with more than 10 independent decisions with around 3 options per decisions. 

Solving problems with larger design spaces will require adopting heuristic search-based approaches that can return good approximations of the set of Pareto-optimal solutions by exploring only small portions of the design space~\cite{harman-sbse}. We intend to implement such search-based approach in the near future. 

We are currently working on extending \radar{}'s modelling language and analysis technique to deal with decision problems with non-mutually exclusive decisions. Such decisions problems have much larger design spaces than the ones we have encountered so far and will require an appropriate heuristic search strategy.

\bibliographystyle{IEEETran}
\bibliography{RadarScalabilityBib}{}

\end{document}